\documentclass[letterpaper,twocolumn,prb,showpacs,floatfix,
superscriptaddress]{revtex4}
\usepackage{graphicx}
\begin{document}

\title{Theory of the Magnetic phase diagram of LiHoF$_4$}

\author{ P. B. Chakraborty}
\affiliation{Department of Physics, Indiana University,
Bloomington, IN 47405} \affiliation{Department of Physics, Yale
University, P. O. Box 208120, New Haven, CT 06520-8120}

\author{P. Henelius}
\affiliation{Condensed Matter Theory, Royal Institute of Technology,
SE-106 91 Stockholm, Sweden}

\author{H. Kj{\o}nsberg}
\affiliation{Telenor R \& D, Snar{\o}yveien 30, Fornebu 1331,
Norway}

\author{A. W. Sandvik}
\affiliation{Department of Physics, Boston University, 590
Commonwealth Ave., Boston, MA 02215} \affiliation{Department of
Physics, {\AA}bo Akademi  University,Porhansgatan 3, FIN-20500,
{\AA}bo , Finland}

\author{S. M. Girvin}
\affiliation{Department of Physics, Yale University, P. O. Box 208120,
  New Haven, CT 06520-8120}

\date{\today}

\begin{abstract}
The properties of LiHoF$_4$ are believed to be well described by a
long-range dipolar Ising model. We go beyond mean-field theory and
calculate the phase diagram of the Ising model in a transverse
field using a quantum Monte Carlo method. The relevant Ising
degrees of freedom are obtained using a non-perturbative
projection onto the low-lying crystal field eigenstates. We
explicitly take the domain structure into account, and the
strength of the near-neighbor exchange interaction is obtained as
a fitting parameter. The on-site hyperfine interaction is
approximately taken into account through a renormalization of the
transverse applied magnetic field. Finally, we propose a
spectroscopy experiment to precisely measure the most important
parameter controlling the location of the phase boundary.

\end{abstract}

\maketitle

%\section{Outline}
%
%\begin{enumerate}
%\item Introduction
%\item Crystal-field hamiltonian
%\begin{enumerate}
%\item Ising system at low temperatures
%\end{enumerate}
%\item Mapping to the Ising system
%\item Computing the phase diagram
%\begin{enumerate}
%\item The effective hamiltonian
%\item The mean-field solution
%\item Domains in LiHoF$_4$
%\item Mean-field theory revisited
%\end{enumerate}
%\item Quantum Monte Carlo simulations on LiHoF$_4$
%\begin{enumerate}
%\item The exchange interaction
%\end{enumerate}
%\item The hyperfine interaction and the final phase diagram
%\item Summary and conclusions
%\item Acknowledgements
%\item Bibliography
%\end{enumerate}

\section{Introduction}

In the last decade, the rare-earth compound LiHoF$_4$ has been
found to display an array of interesting magnetic phenomena. At
high temperatures LiHoF$_4$ is a paramagnet, but there is a
second-order transition to a ferromagnetic state at 1.53
K.\cite{rose91} This Ising magnetic transition is driven by the
weak magnetic dipole interaction, with a strength of order 1K, and
not the more usual Coulomb exchange interaction. The critical
temperature can be lowered by application of a magnetic field
transverse to the easy-axis direction of ferromagnetic ordering.
The magnetic field introduces quantum fluctuations of the spins
and beyond a critical value of $\sim 4.9$ Tesla, destroys
long-range ordering even at zero temperature. LiHoF$_4$ thus
represents a model magnet for studying quantum phase
transitions.\cite{rose96} Since the rate of quantum tunnelling
between different spin configurations can be carefully controlled
with the transverse magnetic field, this material constitutes a
good testing ground for the efficiency of quantum
annealing.\cite{broo99} By substituting the magnetic Ho$^{3+}$
ions with non-magnetic Y$^{3+}$ ions, disorder can be introduced,
and spin-glass behavior has been observed when the magnetic ions
are sufficiently dilute.\cite{rose90} On further dilution the
range of dynamic time scales displays a remarkable narrowing in
what has been called the ``antiglass'' phase.\cite{ghos02}

The magnetic properties of LiHoF$_4$ originate in the Ho$^{3+}$
ions. The ground state of the Ho$^{3+}$ ion in the crystal field
is a doublet, and the first excited state is $\sim 11$K above the
ground state. The crystal field states of the Ho$^{3+}$ ion are
such that there are no matrix elements of the transverse angular
momentum ($J_x,J_y$) within the ground state doublet.  Hence the
transverse susceptibility vanishes (to lowest order in the the
applied field) giving rise to strong Ising anisotropy. Therefore
LiHoF$_4$ in a transverse magnetic field, and at temperatures
lower than $\sim 11$K, is believed to be a very good realization
of a dipolar Ising model
\begin{equation}
H=\frac{1}{2}\sum_{i\ne j}J\frac{r_{ij}^2-3z_{ij}^2}{r_{ij}^5}
S_i^zS_j^z -h^x\sum_i S_i^x, \label{dipole}
\end{equation}
where $J$ is the coupling constant, $r_{ij}$ the interspin
distance and $z_{ij}$ the component of the interspin distance
along the Ising axis. The effective transverse field parameter
$h^x$ is a measure of the (higher order) mixing effects introduced
by the physical transverse magnetic field. The summation is done
over all Ho$^{3+}$ ions, which sit on a body-centered tetragonal
lattice with four Ho$^{3+}$ ions per unit cell.\cite{rose90} %SMG

The goal of the present study is to determine the quantitative
phase diagram of pure LiHoF$_4$ from (quasi-) first principles,
starting from a crystal-field hamiltonian that has been fit to
spectroscopic data.  Bitko et al.\cite{rose96} successfully fit
their phase diagram data using a mean-field theory with two free
parameters, a transverse susceptibility $g_\perp\approx 0.74$ to
replace the Land\'e g factor $g_{\rm L} = 1.25$ which thus
rescales the transverse field, and an effective dipole coupling
strength $J_0$ which rescales the temperature.  This calculation
also did not take into account the domain structure of the
ferromagnetic state.  In another calculation, R{\o}nnow et
al.\cite{jens01} have recently used an RPA method to find the
collective mode softening seen in their neutron scattering
measurements and obtain an estimate of the phase diagram.
%In
%contrast to the thermodynamic measurements of Bitko et
%al.\cite{rose96}, the neutron experiments found no evidence for an
%upturn in the critical field at low temperatures associated with
%the nuclear hyperfine coupling.

%Mean-field theory can, in fact, be used successfully to obtain a
%qualitative agreement with experiments, but it falls far short of
%any reasonable quantitative accuracy. Also, all earlier attempts
%used the full spectrum of crystal-field states to obtain the phase
%diagram, and the very important equivalence of the system to a
%quantum Ising model was entirely unutilised.

%In the present study, we employ a non-perturbative technique to
%map the 17 state crystal field system to an effective Ising model,
%and use quantum Monte Carlo simulations on this Ising model to
%obtain a phase diagram.

The advancements reported in the present study over earlier
attempts are twofold. We develop an effective low-energy
hamiltonian (derived non-perturbatively by projection from the
full 17 state crystal field hamiltonian) that acts on
spin-$\frac{1}{2}$ Ising degrees of freedom, and we go beyond
simple mean-field theory by using extensive quantum Monte Carlo
simulations. An earlier classical Monte Carlo study found a
critical temperature of 1.89 K by extrapolation from rather
limited system sizes.\cite{jens89} However, they did not take into
account a near-neighbor exchange interaction among the Ho$^{3+}$
ions and the structure of domains observed in LiHoF$_{4}$. Yet
another Monte Carlo study finds a transition temperature of 1.51
K, but only by adjusting the strength of the dipolar interaction
to reproduce the experimentally determined ground-state
energy.\cite{xu92} Here we explicitly take the domain structure of
LiHoF$_4$ into account and use a much improved method which vastly
reduces finite-size corrections. We also use a recently introduced
cluster algorithm.\cite{sand03} In this manner we can determine
both the critical temperature and critical field of the effective
Ising model with high precision. This precision is high enough
that the (considerable) uncertainties in the crystal field
parameters (described below) are now the limiting factors
controlling uncertainties in the predicted phase diagram. We will
propose a simple microwave spectroscopy experiment to eliminate
these experimental uncertainties.

\section{The Crystal Field Hamiltonian}

A single Ho$^{3+}$ ion in the crystal LiHoF$_4$ has a partially
filled outermost shell $4f^{10}$, and the ground state electronic
configuration of the Ho$^{3+}$ ion is $^5I_8$ as dictated by
Hund's Rules. The lowest excited electronic configuration of the
ion, $^5I_7$, lies approximately 7400 K above the ground state
configuration, as seen in spectroscopic experiments on
LiHoF$_4$.\cite{chri79} In the range of temperatures of interest
in this article, any configuration mixing of the ground
configuration with the excited ones can thus be safely neglected.
The configuration-mixing due to the crystal field is also assumed
to be small.

Considering only the spin-orbit interaction and the Hund's Rules,
the ground state configuration of the Ho$^{3+}$ ion in LiHoF$_4$
will be (2J+1=)17-fold degenerate. But the interaction of
Ho$^{3+}$ with the Li$^+$ and F$^-$ ions can be captured concisely
in a crystal-field hamiltonian (V$_C$) that lifts the degeneracy
while taking into account the symmetry of the crystal. The
LiHoF$_4$ crystal has S$_4$ symmetry, which partially splits the
17-fold degeneracy. In S$_4$ symmetry, the states of a
configuration with an even number of electrons transform according
to four one-dimensional representations, two of which are related
by time-reversal symmetry. The ground state of the crystal-field
hamiltonian is thus a doublet, belonging to the two related
representations mentioned above, giving rise to a non-Kramers
degenerate ground state. The crystal field hamiltonian depends in
a complicated way on the positions of the various ions inside a
unit cell, but it turns out that one can express V$_C$ in terms of
the total angular momentum ($\vec{J}$) operators of the Ho$^{3+}$
ions by using a set of Steven's equivalent operators and
corresponding phenomenological constants called crystal field
parameters(CFP) which are determined by fitting to experimental
spectroscopic and susceptibility data.\cite{{chri79},{hans75}}(See
Appendix for details on the crystal field hamiltonian).

\subsection{Ising system at low temperatures}
Diagonalizing V$_C$ shows that the lowest excited state in the
spectrum is a singlet, lying $\sim$ 11 K above the ground state
doublet. At temperatures much lower than this gap, only the ground
state doublet can be expected to be significantly populated, and
the low temperature physics can be captured by considering a
two-state system. The two degenerate states (denoted by
$|\alpha\rangle$ and $|\beta\rangle$) can be chosen such that
$\langle\alpha|J^z|\alpha\rangle=-\langle\beta|J^z|\beta\rangle$
and $\langle\alpha|J^z|\beta\rangle=0$. It turns out that the
transverse angular momentum operators J$^x$ and J$^y$ have no
non-zero matrix elements in the degenerate ground-state subspace.
This is the source of the strong Ising anisotropy which causes the
linear susceptibility to vanish in the transverse directions. We
thus identify the two degenerate states as $|\uparrow\rangle$ and
$|\downarrow\rangle$, and it can be expected that the low
temperature physics will be described by an effective Ising model
with spin-$\frac{1}{2}$ degrees of freedom.

The S$_4$ symmetry of the crystal defines an easy axis for
ferromagnetic ordering in the pure LiHoF$_4$ crystal.\cite{rose96}
In the absence of any externally applied magnetic field, the
magnetic dipole interaction among the Ho$^{3+}$ magnetic moments
causes them to align along the c-axis of the unit cell (the
z-direction in this analysis) below 1.53 K and the dipolar Ising
model serves as an adequate effective model for the system.

The situation becomes more subtle when the crystal is subjected to
an external magnetic field perpendicular to the above mentioned
easy axis.\cite{rose96} The magnetic moments couple to the
transverse field through the Zeeman interaction. Restricted within
the ground state configuration J=8, the Wigner-Eckert theorem
yields a Land$\acute{e}$ g-factor $g_L=\frac{5}{4}$ and the Zeeman
term in the hamiltonian can be written as
\begin{equation}
H_Z=-g_L\mu_B\vec{B}\cdot\vec{J}, \label{Zeeman}
\end{equation}
with $\mu_B=0.6717$K/T being the Bohr magneton and
$\vec{B}=B_x\hat{e}_x$.  Because the Zeeman term has no matrix
elements within the two dimensional subspace, it cannot flip a
spin to first order in $\vec{B}$. Thus to see any effect of the
transverse field one must resort to second-order perturbation
theory. Denoting the singlet excited state at $\Delta=$11 K by
$|\gamma\rangle$, one can except to see an effect $\sim$
$(g_L\mu_B)^2\frac{B_x^2}{\Delta}|\langle\gamma|J^x|\alpha,\beta\rangle|^2$
on the energies of the ground states. A naive application of
degenerate second-order perturbation theory thus suggests that the
effect of the transverse field should be proportional to $B_x^2$.

This perturbation theory scenario breaks down in the quantum
critical regime as can be seen in a comparison of the energies. If
we assume that the magnetic moment is fully polarized in the
transverse direction ($\langle J^x\rangle=J=8$) at the quantum
critical point of $B_x^c=4.9$T at T=0,\cite{rose96} a simple
estimate of the Zeeman energy is given by
\begin{equation}
E_Z=g_L\mu_BB_x^c\langle J^x\rangle=32.91\rm{K}, \label{estimate}
\end{equation}
significantly larger than $\Delta$. This demonstrates that the
mixing of the ground-state doublet with all the higher-lying
states must be considered at large transverse fields and
second-order perturbation theory is not sufficient to incorporate
the effect of $\vec{B}$ in the quantum critical regime. We
describe below a non-perturbative scheme to capture this physics.

\section{Mapping to the Ising system }

The magnetic properties of LiHoF$_4$ are determined by three kinds
of interactions: a long-range magnetic dipole interaction among
the Ho$^{3+}$ magnetic moments, a near-neighbor exchange
interaction which we assume to be small and isotropic, and an
isotropic hyperfine interaction between the electronic and nuclear
magnetic moments on the same site. Therefore, the complete
hamiltonian of a LiHoF$_4$ crystal in a transverse magnetic field
can be written as
\begin{eqnarray}
H&=&\sum_iV_C(\vec{J}_i) - g_L\mu_B\sum_iB_{x}J_i^x\nonumber\\
&+&\frac{1}{2}(g_L\mu_B)^2\sum_{i\ne
j}\mathcal{L}_{ij}^{\mu\nu}J^{\mu}_iJ^{\nu}_j +
\frac{1}{2}(g_L\mu_B)^2\frac{J_{\rm{ex}}}{a^3}\sum_{i,nn}\vec{J_i}.
\vec{J}_{nn}\nonumber\\ &+&A\sum_i(\vec{I}_i.\vec{J}_i),
\label{firstprin}
\end{eqnarray}
where $\mu,\nu=x,y,z$. $\mathcal{L}_{ij}^{\mu\nu}$ contains the
position dependence of the magnetic dipole interaction so that
\begin{equation}
\mathcal{L}_{ij}^{\mu\nu}=\frac{\delta^{\mu\nu}|\vec{r}_{ij}|^2 -
3(\vec{r}_{ij})^{\mu}(\vec{r}_{ij})^{\nu}}{|\vec{r}_{ij}|^5}.
\label{dipole2}
\end{equation}
J$_{\rm{ex}}$ is the dimensionless strength of the
antiferromagnetic exchange interaction (whose strength is unknown
at this point), $nn$ signifies that the sum is to be carried out
over nearest neighbors only and $a$(=5.175$\mathring{A}$) is the
length of the unit cell in the x-y plane. A is the strength of the
hyperfine interaction(0.039 K) and $\vec{I}_{i}$ is the total
angular momentum vector of the Ho-nucleus at the i-th site,
$I=\frac{7}{2}$.

\begin{figure}[htp]
\resizebox{\hsize}{!}{\includegraphics[clip=true]{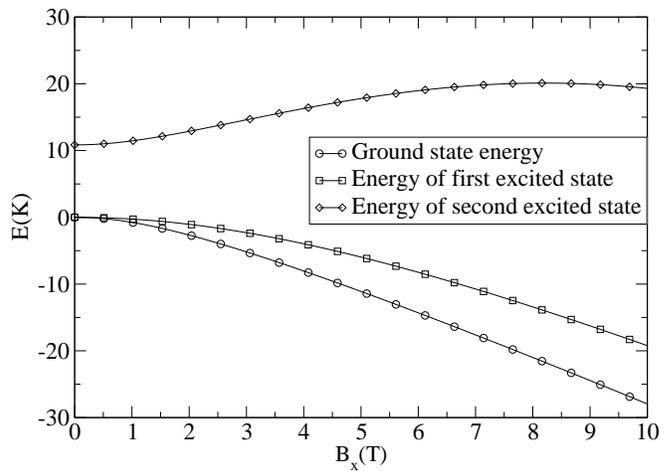}}
\caption{\label{fig:energy} Lifting of ground state degeneracy by the
transverse field. }
\end{figure}

To reduce the hamiltonian that acts on a
(2J+1)$\times$(2I+1)=136-dimensional Hilbert space to an effective
Ising model with spin-$\frac{1}{2}$ degrees of freedom, we neglect
the relatively weak hyperfine interaction in a first
approximation. If $A=0$, the only relevant degrees of freedom are
the electronic angular momenta $\vec{J}$, and, for $J=8$, the
single-site Hilbert space is 17-dimensional. The transverse field
splits the degeneracy in the ground state subspace and also mixes
the 15 higher-lying crystal-field states with the two lowest
states. The single-site hamiltonian (neglecting the hyperfine
interaction)
\begin{equation}
H_{T}=V_{C}(\vec{J}) - g_{L}\mu_{B}B_{x}J^{x}
\label{truncate}
\end{equation}
is diagonalized numerically for all values of the transverse
field. For a given B$_{x}$, let the two lowest states be denoted
by $|\alpha(B_{x})\rangle$ and $|\beta(B_{x})\rangle$ and their
energies be denoted by $E_{\alpha}(B_{x})$ and $E_{\beta}(B_{x})$.
Then in a two-dimensional Hilbert space which is spanned by
$|\alpha(B_{x})\rangle$ and $|\beta(B_{x})\rangle$ (identified as
$|\rightarrow\rangle$ and $|\leftarrow\rangle$, respectively)
$H_{T}$ can be written as
\begin{equation}
H_{T}=E_{CM}(B_{x}) - \frac{1}{2}\Delta(B_{x})\sigma^{x},
\label{splitting}
\end{equation}
where $E_{CM}(B_{x})=\frac{1}{2}(E_{\alpha}(B_{x}) +
E_{\beta}(B_{x}))$ and $\Delta(B_{x})=E_{\beta}(B_{x}) -
E_{\alpha}(B_{x})$.  Thus we see that the energy difference
between the degenerate states caused by the transverse field can
already be interpreted as an effective magnetic field acting on
spin-$\frac{1}{2}$ degrees of freedom at each site.
Fig.~(\ref{fig:energy}) demonstrates how the transverse field
continuously splits the degeneracy between the two ground states.

To generate the complete first-principles hamiltonian in
Eq.~(\ref{firstprin}) in terms of the Ising spins, one must
project the $\vec{J}$-operators onto the two dimensional subspace.
Recognizing the fact that every two dimensional hermitian matrix
can be uniquely expanded in terms of the two dimensional unit
matrix and the three Pauli matrices, we evaluate the matrix
elements of the $\vec{J}$-operators within the two dimensional
subspace for each value of the magnetic field and obtain their
representations in terms of spin-$\frac{1}{2}$ operators. Every
operator $J^{\mu}$ ($\mu$=x, y, z) can be written as
\begin{equation}
J^{\mu}=C_{\mu} + \sum_{\nu=x,y,z}C_{\mu\nu}(B_{x})\sigma^{\nu}.
\label{map1}
\end{equation}
The Pauli matrices $\sigma^{\nu}$ requires the choice of a basis
$|\uparrow\rangle$ and $|\downarrow\rangle$ in terms of the states
$|\alpha\rangle$ and $|\beta\rangle$. We choose the basis vectors
such that $|\uparrow\rangle = \frac{1}{\sqrt{2}}(|\alpha\rangle +
\exp(i\theta)|\beta\rangle)$ and $|\downarrow\rangle =
\frac{1}{\sqrt{2}}(|\alpha\rangle - \exp(i\theta)|\beta\rangle)$.
The phase $\theta$ is chosen such that the matrix elements of the
operator $J^{z}$ are real. In this basis, we have
\begin{equation}
J^{z}_{i}=C_{zz}\sigma^{z}_{i}. \label{zmap1}
\end{equation}
The operators $J^{x}$ and $J^{y}$ are also projected into this
basis.
 The following table presents the values of the various
parameters $C_{\mu}$ and $C_{\mu\nu}$, for B$_{x}$=4.9 T, which is
the experimentally observed critical transverse field in the limit
of vanishing temperature.

\begin{table}[htp]
\begin{ruledtabular}
\begin{tabular}{c|c|c|c|c|c}
C$_{z}$ & 0 & C$_{x}$ & 3.12 & C$_{y}$ & 0.01\\\hline C$_{zx}$ & 0
& C$_{xx}$ & 0.60 & C$_{yx}$ & -0.09\\\hline C$_{zy}$ & 0 &
C$_{xy}$ & 0.01 & C$_{yy}$ & 1.05\\\hline C$_{zz}$ & 5.14 &
C$_{xz}$ & 0 & C$_{yz}$ & 0
\end{tabular}
\end{ruledtabular}
\caption{The strengths of various coefficients C$_{\mu\nu}$ for B$_{x}$=4.9 T.}
\label{tab1}
\end{table}
We replace the $\vec{J}$-operators by their corresponding
representations in the two-dimensional basis in the hamiltonian in
Eq.~(\ref{firstprin}) (The hyperfine interaction is neglected at
this stage). This leads to an extremely complicated hamiltonian
that acts on the Ising subspace. The projection generates various
new kinds of interactions among the effective Ising spins. For
example, the term
$$H_{zx}=\frac{1}{2}(g_{L}\mu_{B})^{2}\sum_{i\neq j}\mathcal{L}^{zx}_{ij}J^{z}_{i}J^{x}_{j}$$
in Eq.~(\ref{firstprin}) gives rise to a set of complicated
interactions in the Ising subspace that are written as
 \begin{eqnarray}
H_{zx}&=&\frac{1}{2}(g_{L}\mu_{B})^{2}\sum_{i\neq
j}\mathcal{L}^{zx}_{ij}(C_{zz}\sigma^{z}_{i})(C_{x} +
C_{xx}\sigma^{x}_{j} +
C_{xy}\sigma^{y}_{j})\nonumber\\
&=&\frac{1}{2}(g_{L}\mu_{B})^{2}\sum_{i}(\sum_{j\neq
0}\mathcal{L}^{zx}_{j0})C_{zz}C_{x}\sigma^{z}_{i}\nonumber\\
& &+\frac{1}{2}(g_{L}\mu_{B})^{2}\sum_{i\neq
j}\mathcal{L}^{zx}_{ij}C_{zz}C_{xx}\sigma^{z}_{i}\sigma^{x}_{j}\nonumber\\
& &+\frac{1}{2}(g_{L}\mu_{B})^{2}\sum_{i\neq
j}\mathcal{L}^{zx}_{ij}C_{zz}C_{xy}\sigma^{z}_{i}\sigma^{y}_{j}.
\label{projectionexample}
\end{eqnarray}
In Eq.~(\ref{projectionexample}), the quantities $C_{zz}$ etc.
depend on the transverse magnetic field, $B_{x}$. The strength of
the effective interactions generated by the Ising projection is
determined not only by the parameters $C_{\mu\nu}$, but also by
the parameters $\mathcal{L}^{\alpha\beta}_{ij}$, that depend on
the inter-spin vector $\vec{r}_{ij}$. Thus we can compare the
strengths of the various effective interactions, and we find that
the largest effective interaction is $J^{z}_{i}J^{z}_{j}\propto
(C_{zz})^{2}\sigma^{z}_{i}\sigma^{z}_{j}$. This interaction is
larger than the other interaction terms by two orders of magnitude
for the entire range of magnetic fields in question (except for
some constant terms that just serves to define a new zero of
energy). Therefore, to an accuracy of $\sim$ 1\%, the effective
Ising model for LiHoF$_4$ can now be written as
\begin{eqnarray}
H_{\rm{Ising}}&=&\sum_{i}E_{CM,i}(B_{x}) -
\frac{1}{2}\Delta(B_{x})\sum_{i}\sigma^{x}_{i}\nonumber\\
&+&\frac{1}{2}(g_{L}\mu_{B}C_{zz})^{2}\sum_{i\ne
j}\mathcal{L}^{zz}_{ij}\sigma^{z}_{i}\sigma^{z}_{j}\nonumber\\
&+&\frac{1}{2}(g_{L}\mu_{B}C_{zz})^{2}\frac{J_{\rm
ex}}{a^{3}}\sum_{i,nn} \sigma^{z}_{i}\sigma^{z}_{nn}.
\label{isingfirst}
\end{eqnarray}
The Ising hamiltonian of Eq.~(\ref{isingfirst}) acts on effective
spin-$\frac{1}{2}$ objects(``pseudospins'') located at the lattice
sites of Ho$^{3+}$ ions in the pure crystal LiHoF$_{4}$, and the
interaction strengths and the effective transverse field depend on
the physical transverse field B$_{x}$ implicitly through the
parameters $C_{zz}$ and $\Delta$. The parameter $E_{CM}$ provides
a B$_{x}$-dependent zero of energy and will be ignored for the
purpose of computing the magnetic phase diagram.

\section{ Computing the phase diagram }

Eq.~(\ref{isingfirst}) is a particular example of a general class
of hamiltonians in which the various terms in the hamiltonian do
not commute with each other, and in the interesting parameter
regime around the quantum critical point, they are comparable in
strength. Therefore, quantum fluctuations can become strong enough
in the system to destroy long-range order even at T=0. An Ising
model of spin-$\frac{1}{2}$ objects on a three dimensional lattice
placed in a magnetic field transverse to the Ising direction is a
very good prototype of such a hamiltonian. We utilize
Eq.~(\ref{isingfirst}) as a starting point to compute the phase
diagram of LiHoF$_4$ in various ways, notably using mean-field
theory and quantum Monte Carlo simulations. As a spin-off, we are
able to calculate the strength of the exchange interaction
J$_{\rm{ex}}$. Finally, we compare our results with existing
experimental data,\cite{rose96} which were obtained from magnetic
susceptibility measurements.

\subsection { The effective hamiltonian }

An unusual feature of the Ising hamiltonian in
Eq.~(\ref{isingfirst}) is that the strength of the dipole
interaction itself seems to depend explicitly on the strength of
the physical transverse field through the parameter C$_{zx}$.
Ordinarily, the interaction term in an Ising model is free of any
dependence on external magnetic fields. The popular hamiltonian
governing the quantum Ising model is generically written as
\begin{equation}
H_{\rm{gen}}=\frac{1}{2}\sum_{i \ne
j}J_{ij}\sigma^{z}_{i}\sigma^{z}_{j} -
h^{x}\sum_{i}\sigma^{x}_{i}, \label{popular}
\end{equation}
where $J_{ij}$ is the pairwise interaction term (dipole
interaction, near-neighbor exchange ..) which usually depends on
the difference in spatial positions of the spins, and $h^{x}$ is
the magnetic field that acts as the source of the quantum
fluctuations. With some appropriate definitions, we are able to
define an effective model H$_{\rm{eff}}$ that is of the same
general form as Eq.~(\ref{popular}).

The reduction to a two dimensional Hilbert space from a 17-dimensional
Hilbert space gives rise to renormalized Land$\acute{\rm{e}}$ g-factor
g$_{||}$, defined as
\begin{equation}
g_{||}=2g_{L}\langle\alpha(B_{x}=0|J^{z})|\alpha(B_{x}=0)\rangle=13.8
. \label{gparallel}
\end{equation}
The factor $2g_{L}C_{zz}(B_{x})$ can be interpreted as a
renormalization of the individual magnetic moments due to the
presence of the transverse magnetic field. To extract this
renormalization factor, we define a dimensionless ratio
$\epsilon(B_{x})$ such that
\begin{equation}
\epsilon(B_{x})=\frac{2g_{L}C_{zz}(B_{x})}{g_{||}}.
\label{epsilon}
\end{equation}
In the classical limit, $B_{x}=0$, $\epsilon(B_{x})=1$ by definition.
This makes it possible to define an effective hamiltonian
H$_{\rm{eff}}$, which is essentially identical to Eq.~(\ref{popular}).
\begin{equation}
H_{\rm{Ising}}=[\epsilon(B_{x})]^{2}H_{\rm{eff}},
\label{effective}
\end{equation}
where
\begin{eqnarray}
H_{\rm{eff}}&=&\frac{1}{2}\sum_{i \ne
j}\left(\frac{g_{||}\mu_{B}}{2}\right)^{2}\mathcal{L}^{zz}_{ij}\sigma^{z}_{i}
\sigma^{z}_{j}\nonumber\\&+&\frac{1}{2}\sum_{i,nn}\left(\frac{g_{||}
\mu_{B}}{2}\right)^{2}\frac{J_{\rm{ex}}}{a^{3}}\sigma^{z}_{i}\sigma^{z}_{nn}
\nonumber\\
&-&\frac{\Delta(B_{x})}{2[\epsilon(B_{x})]^{2}}\sum_{i}\sigma^{x}_{i}.
\label{heff}
\end{eqnarray}
We define every length in the system in units of $a=5.175
\mathring{A}$, such that
$\tilde{r}^{\mu}_{ij}=\frac{r^{\mu}_{ij}}{a}$. Then the spatial
dependence of the dipole interactions can be expressed in the
terms of dimensionless quantities $\rm{L}^{\mu\nu}_{ij}$, defined
as $\mathcal{L}^{\mu\nu}_{ij}=\frac{\rm{L}^{\mu\nu}_{ij}}{a^{3}}$.
Replacing the constants in Eq.~(\ref{heff}), we see that
$\left(\frac{g_{||}\mu_{B}}{2}\right)^{2}\frac{1}{a^{3}}=0.214$ K,
and $\left(\frac{g_{||}\mu_{B}}{2}\right)=4.635$ K/T. These
constants specific to LiHoF$_4$ are substituted in
Eq.~(\ref{heff}), and we obtain
\begin{eqnarray}
H_{\rm{eff}}&=&\frac{1}{2}\times0.214\sum_{i \ne
j}{\rm{L}}^{zz}_{ij}\sigma^{z}_{i}\sigma^{z}_{j}\nonumber\\
&+&\frac{1}{2}\times0.214\sum_{i,nn}J_{\rm{ex}}\sigma^{z}_{i}\sigma^{z}_{j}
\nonumber\\&-&4.635B^{x}_{\rm{eff}}\sum_{i}\sigma^{x}_{i}.
\label{effham}
\end{eqnarray}
A comparison of Eq.~(\ref{heff}) and Eq.~(\ref{effham}) immediately
yields the crucial correspondence between the effective transverse
magnetic field $B^{x}_{\rm{eff}}$ (expressed in Tesla) and the
physical transverse field $B_{x}$ (also expressed in Tesla).
\begin{equation}
B^{x}_{\rm{eff}}=\frac{\Delta(B_{x})}{2\times4.635\times[\epsilon(B_{x})]^{2}}.
\label{magnetmap}
\end{equation}

\begin{figure}[htp]
\resizebox{\hsize}{!}{\includegraphics[clip=true]{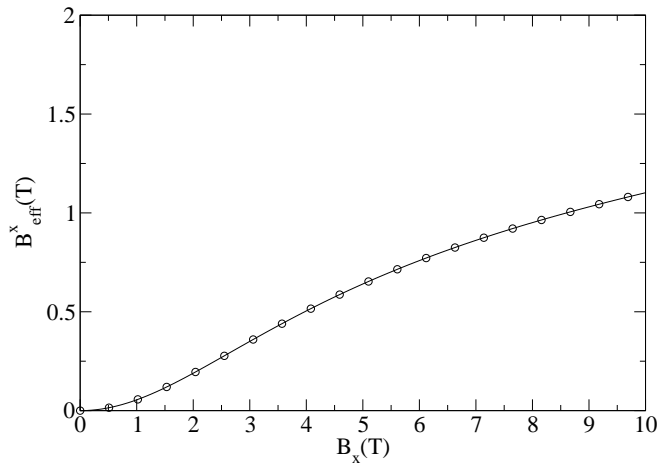}}
\caption{\label{fig:efffield} The effective magnetic field. }
\end{figure}

We note that the dependence of $B^{x}_{\rm{eff}}$ on $B_{x}$ in
Fig.~(\ref{fig:efffield}) is indeed quadratic for small values of
$B_{x}$, as dictated by second-order perturbation theory, but the
dependence becomes increasingly weaker as $B_{x}$ increases. This
demonstrates that in the quantum critical regime of large
transverse fields and vanishing temperature, perturbation theory
arguments are not adequate.

At this point, a subtlety in the definition of ``temperature'' needs
to be mentioned. All our further calculations use $H_{\rm{eff}}$ as in
Eq.~(\ref{effham}), but the physical hamiltonian is $H_{\rm{Ising}}$,
which modifies the energy scale by a factor $[\epsilon(B_{x})]^{2}$
(See Eq.~(\ref{effective})). In both mean-field theory and quantum
Monte Carlo calculations, the temperature enters only in the
definition of the Boltzmann weight of a configuration or a
state-vector in the form (k$_{\rm{B}}$=1)
\begin{eqnarray}
\rho&=&\exp\left(-\frac{H_{\rm{Ising}}}{T}\right)\nonumber\\
&=&\exp\left(-\frac{[\epsilon(B_{x})]^{2}H_{\rm{eff}}}{T}\right)\nonumber\\
&=&\exp\left(-\frac{H_{\rm{eff}}}{T_{\rm{eff}}}\right),
\label{tempdef}
\end{eqnarray}
where the physical temperature $T$ is related to the effective
temperature $T_{\rm{eff}}$ by
\begin{equation}
T=[\epsilon(B_{x})]^{2}T_{\rm{eff}} . \label{teff}
\end{equation}

\subsection{A possible three-state description of the effective
Ising model}

In this section, we digress from the discussion a bit to show that
it is also possible to capture the degeneracy splitting
$\Delta(B_{x})$ using a simple 3-state model of the system, but it
leaves out the effect of the matrix elements of the ground state
doublet with the higher-lying states. If we consider the states
$|\alpha\rangle$,$|\beta\rangle$ and $|\gamma\rangle$ at $B_{x}=0$
as the basis states, then the crystal field hamiltonian $V_{C}$ is
simply diagonal in this basis, and looks like
\begin{equation}
V_{C}=\left(%
\begin{array}{ccc}
  0.0 & 0 & 0 \\
  0 & 0.0 & 0 \\
  0 & 0 & \Delta \\
\end{array}%
\right), \label{threevc}
\end{equation}
where $\Delta=10.8\rm{K}$.

The operator $J^{x}$ has only two non-zero matrix elements
restricted within the three states in the absence of a transverse
field. By choosing the phase of the wave-functions properly, it is
possible to make both of them real. These matrix elements are
\begin{equation}
\langle\alpha|J^{x}|\gamma\rangle=\langle\beta|J^{x}|\gamma\rangle=\rho
, \label{threestatematrels}
\end{equation}
where $\rho=2.4$. Then the Hamiltonian in Eq.~(\ref{truncate}) can
be written in the three-state basis as
\begin{equation}
H_{T}=\left(%
\begin{array}{ccc}
  0& 0 & -g_{L}\mu_{B}B_{x}\rho \\
  0 & 0 & -g_{L}\mu_{B}B_{x}\rho \\
  -g_{L}\mu_{B}B_{x}\rho & -g_{L}\mu_{B}B_{x}\rho & \Delta \\
\end{array}%
\right).
\end{equation}
This hamiltonian can be exactly diagonalized, and defining
$\tilde{B_{x}}=g_{L}\mu_{B}B_{x}\rho$, the eigenvalues are
\begin{eqnarray}
  E_{0} & = & \frac{\Delta}{2}-\frac{1}{2}\sqrt{\Delta^{2}+8\tilde{B_{x}^{2}}}
  \\\nonumber
  E_{1} & =& 0 \\\nonumber
  E_{2} & = &
  \frac{\Delta}{2}+\frac{1}{2}\sqrt{\Delta^{2}+8\tilde{B_{x}^{2}}}.
\label{threeeigenvalues}
\end{eqnarray}
The eigenstates can be schematically written as
\begin{eqnarray}
  |\psi_{0}\rangle & = & \frac{1}{\sqrt{2}}(|\alpha\rangle+|\beta\rangle)+\varepsilon_{0}|\gamma\rangle
  \\\nonumber
  |\psi_{1}\rangle & = & \frac{1}{\sqrt{2}}(|\alpha\rangle-|\beta\rangle)
  \\\nonumber
   |\psi_{2}\rangle& = & |\gamma\rangle +
   \varepsilon_{2}(|\alpha\rangle+|\beta\rangle),
\label{threeeigenstates}
\end{eqnarray}
where the symbols $\varepsilon_{0}$ and $\varepsilon_{2}$ signify
small admixtures.

It is readily seen that the transverse field splits the degeneracy
by an amount $\Delta(B_{x})=E_{1}-E_{0}$. The gap is quadratic in
$B_{x}$ when $\tilde{B}_{x}\ll\Delta$, but it is, in fact, linear
when $\tilde{B}_{x}\gg\Delta$. However, we see that the lowest
excited state has an eigenvalue of zero for all $B_{x}$, unlike
the real system, where it bends downwards with increasing $B_x$.
Thus the three state model cannot capture the repulsion from the
higher-lying states. Even then, it explicitly shows that the
physics near the quantum critical point is ill-described by
second-order perturbation theory.

\subsection{Domains in LiHoF$_{4}$ and the mean-field solution}

The magnetic dipole interaction is long-range, falling off as
$\frac{1}{r^{3}}$, where $r$ is the distance between two spins. A
dipole-coupled Ising ferromagnet is expected to be well explained
by mean-field theory for $d \geq 3$. As we shall see, mean-field
theory does capture the qualitative behavior in the phase diagram,
but is not very reliable for making quantitative predictions.

Our starting point is Eq.~(\ref{effham}). The individual
spin-$\frac{1}{2}$ degrees of freedom sit on a tetragonal unit
cell of dimensions $(a,a,c)=(1,1,2.077)$ in units of $a=5.175
\mathring{A}$. Each unit cell has four spins, at the positions we
denote as
$[(0,0,0),(0,\frac{a}{2},\frac{c}{4}),(\frac{a}{2},\frac{a}{2},-\frac{c}{2})
\hspace{1mm} {\rm{and}}
\hspace{1mm}(\frac{a}{2},0,-\frac{c}{4})]$. All the positions in
the unit cell are measured from the Bravais lattice vector denoted
by $\vec{R} + (0,0,0)$.

The long-range nature and angular dependence of the dipole
interaction create a complex model in which the actual shape and
lattice of the sample influences the ground state. Spins that lie
in the same ab-plane want to antialign, while spins along the same
c-axis want to order ferromagnetically. This angular dependence
tends to favor ferromagnetism in long thin samples, while it
prohibits a ferromagnetic state in spherical samples. Luttinger
and Tisza\cite{lutt46} found, for example, a ferromagnetic ground
state for the face-centered lattice in the shape of a prolate
spheroidal sample of axis ratio larger than 6. However, in
experiments on LiHoF$_4$ spherical,\cite{rose96}
cubic\cite{rose90} and rectangular\cite{rose91} shapes have been
used with no apparent dependence of the results on the shape of
the sample.

The reason for this sample-shape independence lies in the domain
structure of LiHoF$_4$. Experimentally, there is
evidence\cite{{cook75},{batt75}} that LiHoF$_4$ forms long
needle-shaped domains, with the long direction being along the
c-axis. Since this is an Ising system, the magnetization
alternates in sign between adjacent domains, which minimizes the
global energy stored in the macroscopic fields outside the sample.
Thus it would be incorrect, in the context of mean-field theory,
to assume that the magnetization is uniform over the entire
sample, since in fact, it is only uniform over a single domain.

To incorporate the effect of domains, we assume that an imaginary
sphere sits deep inside a single domain.\cite{heid00} A single
spin at the center of this sphere now experiences an effective
field from the orientation of the magnetic dipoles both inside and
outside the sphere. In the ferromagnetically ordered state, the
magnetization of the domain in question can be assumed to be
uniform, denoted now by $M^{z}$. The magnetic field acting on the
domain as a whole is the external field $B^{z}_{\rm{ext}}$, and
the susceptibility $\chi$ of the domain can be found from the
relation
\begin{equation}
M^{z}=\chi B^{z}_{\rm{ext}}. \label{chidomain}
\end{equation}

On the other hand, if we consider the small imaginary sphere deep
inside the domain, the susceptibility $\chi_{\rm{sph}}$ of the
sphere can be found from the equation relating the magnetization
inside the sphere (since the sphere is a part of the domain, the
magnetization inside it is $M^{z}$) and the magnetic field acting
on the sphere, $B^{z}_{\rm{sph}}$,
\begin{equation}
M^{z}=\chi_{\rm{sph}} B^{z}_{\rm{sph}}. \label{chisphere}
\end{equation}

We treat the magnetic field produced by the spins {\em{outside}}
the sphere using mean-field theory. The spins at the surface of
the hollow sphere produces a polarization field proportional to
$M^{z}$, while the spins at the outer surface of the domain
produces a demagnetization field, also proportional to $M^{z}$,
that is characterized by a demagnetization factor $N^{z}$,
dependent on the {\em{shape}} of the domain. Therefore, we can
write down the general relation\cite{jack}
\begin{equation}
B^{z}_{\rm{sph}}=B^{z}_{\rm{ext}} + \frac{8\pi}{3}M^{z} -
N^{z}M^{z}. \label{domainfield}
\end{equation}

For a thin needle-shaped domain, $N^{z}=\frac{4\pi}{3}$, and
substituting Eq.~(\ref{domainfield}) in Eqs.~(\ref{chidomain}) and
(\ref{chisphere}) leads to the following relation
\begin{equation}
\chi=\frac{1}{\chi^{-1}_{\rm{sph}} - \frac{4\pi}{3}}.
\label{chirelation}
\end{equation}

Eq.~(\ref{chirelation}) above is the magnetic version of the
Clausius-Mosotti relation,\cite{jack} which relates the
macroscopic electric susceptibility of a system with the
microscopic molecular polarizability. In fact, it is in some ways
easier to derive from the analogy with electric dipoles. In that
case, the field produced by the \emph{charge density} at the
surface of the spherical cavity is $\frac{4\pi}{3}M^{z}$ while the
small random patches of \emph{charge} where the domains reach the
surface of the sample produce negligible field.

 We emphasize that this relation is obtained from a single assumption that the spins
{\em{outside}} the sphere are treated in mean-field fashion. The
microscopic variable, $\chi_{\rm{sph}}$, can be calculated
approximately, using mean-field theory, or exactly, using quantum
Monte Carlo(QMC) simulations over a sphere. The next part of this
section explores how the domain structure can be incorporated in
the mean-field scenario, while the following section contains the
results from QMC simulations.

At this point, a comment on the effect of the transverse magnetic
field on the domain formation is necessary. From the mapping to the
Ising model, it is seen that the contribution of the transverse dipole
interaction, ${\rm{L}}_{ij}^{xx}$ is negligible compared to the
longitudinal dipole interaction, ${\rm{L}}_{ij}^{zz}$. Therefore, we
may assume that the transverse magnetic field polarizes the Ising
spins along the x-direction uniformly throughout the sample. In other
words, the x-component of the magnetization, $M^{x}$, is unaware of
the domain structure formation in LiHoF$_{4}$.

\subsection{ Mean-Field theory}

Let us now calculate the mean-field critical temperature for a
long, needle-shaped domain. We therefore place the sphere deep
inside a single domain. The mean-field critical temperature is
given by the local field, which is given by the sum over all
dipoles within the sphere. In the presence of domains, this field
must be augmented by the field acting on the sphere,
$B^{z}_{\rm{sph}}$.

With no domains a single spin at the center of the sphere therefore
experiences an effective longitudinal mean-field given by
\begin{equation}
B^{z}_{\rm{MF}}=\left(\sum_{j \ne
0}^{\rm{sphere}}{\rm{L}}^{zz}_{0j}\right)\times
\left(\frac{m^z}{a^{3}}\right),
\label{meanfieldnodomain}
\end{equation}
where $m^z$ is the magnetic moment of a single dipole,
\begin{equation}
m^{z}=\left(\frac{g_{||}\mu_{B}}{2}\right)\langle\sigma^{z}\rangle.
\label{magdenop}
\end{equation}
The value of the lattice sum in Eq.~(\ref{meanfieldnodomain}) can
be easily obtained by summing over larger and larger spheres, and
the convergence is found to be very rapid. The value for LiHoF$_4$
is
\begin{equation}
\sum_{j\ne 0}^{\rm{sphere}}\rm{L}^{zz}_{0j}=3.205 .
\label{latticesum}
\end{equation}

 In order to take the domain into account we combine
Eqs.~(\ref{domainfield}) and Eq.~(\ref{meanfieldnodomain}). Since
the macroscopic magnetization is given by
\begin{equation}
M^{z}=N_{0}m^{z}, \label{locmagden}
\end{equation}
where $N_{0}$ is the number density of dipoles, we simply have to
augment the value of the lattice sum in Eq.~(\ref{latticesum}) by
the value $\frac{4\pi}{3}N_{0}a^{3}$. Since $N_{0}=\frac{4}{a
\times a \times c}$ and $c/a=2.077$, it follows that
$\frac{4\pi}{3}N_{0}a^{3}=8.067$.

This leads to the value of the critical temperature at the
classical limit.
\begin{equation}
T_{\rm{C}}=0.214 \rm{K} \times (3.205 + 8.067) \approx 2.41 \rm{K}
. \label{tcwithdomain}
\end{equation}

On the other hand, the new critical effective field at the quantum
limit is given by
\begin{equation}
B^{x}_{{\rm{eff}},{\rm{C}}}=\frac{2.41 \rm{K}}{4.635 \rm{K}/\rm{T}
} \approx 0.52 \rm{T}, \label{bxcwithdomains}
\end{equation}
which corresponds to a physical critical transverse field of
$B_{x,{\rm{C}}}\approx 4.11 \rm{T}$.

We note that the critical temperature at the classical limit, 2.41
K, is 57\% larger than the experimentally observed value of 1.53
K. This behavior is encouraging, in the sense that mean-field
theory should always overestimate the ordered region. As far as
the quantum limit is concerned, it may seem that the mean-field
estimate is wrong, since 4.11 T is about 16\% smaller than the
experimentally observed value of 4.9 T. However, we have entirely
neglected the on-site hyperfine interaction between the electronic
cloud and the Ho-nucleus. At low temperatures, this interaction
affects the phase boundary significantly.\cite{rose96} Therefore,
at this point, we cannot yet compare our results for the quantum
limit with experiment.

As a consistency check to our mean-field treatment of the domain
structure, we perform the lattice sum directly over a long
cylinder. The summation is done by considering a series of long
cylinders with increasing base area. For each cylinder with a fixed
base area, the length is increased until convergence is obtained. The
series converges quite rapidly and as a final result we have
\begin{eqnarray}
-\sum_{j \ne
  0}^{\rm{cylinder}}\left(\frac{{\tilde{r}}_{j}^{2}-3{\tilde{z}}_{j}^{2}}{{
  \tilde{r}}_{j}^{5}}\right)&=&11.272\nonumber\\ &=&3.205 + 8.067
  ,
\label{cylinder}
\end{eqnarray}
in perfect agreement with the mean-field approach described above.
We note that the mean-field theory calculation in Ref.[2] does not
take into account the physics of domain formation but rather
simply re-scales the effective couplings to force a fit to the
observed $T_{C}$ and $B_{x,C}$.

\section{Quantum Monte Carlo simulations on LiHoF$_4$ }

We perform extensive QMC simulations of the dipole-coupled quantum
Ising model, described by the hamiltonian in Eq.~(\ref{effham}).
We use a recently introduced\cite{sand03} stochastic series
expansion (SSE) cluster quantum Monte Carlo method for which
computational time scales as $N \ln(N)$, where the number of spins
is given by $N$. In traditional single-spin-flip simulations of
long-range models the computational time typically scales as
$N^{2}$, due to the summation over interactions between all pairs
of spins. The improved scaling enables us to increase the
precision and reach large system sizes.

Our starting point is again Eq.~(\ref{chirelation}), which relates a
microscopic quantity, $\chi_{\rm{sph}}$, to a macroscopic quantity
$\chi$. From Eq.~(\ref{chirelation}), we find that $\chi$ can diverge,
even when $B^{z}_{\rm{ext}}$ vanishes, when
\begin{equation}
\chi_{\rm{sph}}=\frac{3}{4\pi}. \label{critcon}
\end{equation}
We proceed to evaluate $\chi_{\rm{sph}}$ exactly using QMC simulations
as a function of $T_{\rm{eff}}$ and $B^{x}_{\rm{eff}}$. At the point
where the condition in Eq.~(\ref{critcon}) is met, $\chi$, the
macroscopic susceptibility, diverges and the system is critical.

Truncating the sum in the hamiltonian at the boundary of the
sphere and treating the dipoles outside the sphere in mean-field
fashion can be considered a boundary condition in the Monte Carlo
simulation. This so-called reaction-field method\cite{bark73} is
not the only way to incorporate the long-range interaction in an
efficient way. An alternative is to consider a large number of
periodic images of the simulation volume. Performing the necessary
sums over all dipoles in the simulation volume and the image
volumes is very time consuming, but the sums can often be
efficiently performed using the Ewald summation
technique.\cite{ewal21} In this work we have not attempted to
compare the relative advantages of the methods, but we have found
the reaction-field method very convenient. An earlier
comparison\cite{kret79} found the Ewald summation method also
quite reliable.

Operationally, the susceptibility $\chi_{\rm{sph}}$ is defined as the
spin-spin correlation function in imaginary time, averaged over all
spins inside the sphere.
\begin{equation}
\chi_{\rm{sph}}=\frac{\alpha}{N}\sum_{ij}\int_{0}^{\beta}d\tau\langle
\sigma^{z}_{i}(\tau)\sigma^{z}_{j}(0)\rangle .\label{corr}
\end{equation}
The prefactor $\alpha$ is given by
\begin{equation}
\alpha=N_{0}\left(\frac{g_{||}\mu_{B}}{2}\right)^{2}.
\label{alpha}
\end{equation}
The imaginary time integral in Eq.~(\ref{corr}) can be evaluated
directly by the SSE method.\cite{sand91} The condition for
criticality can be now be written as
\begin{eqnarray}
\frac{1}{N}\sum_{ij}\int_{0}^{\beta}d\tau\langle\sigma^{z}_{i}(\tau)
\sigma^{z}_{j}(0)\rangle&=&\frac{3}{4\pi
N_{0}}\left(\frac{2}{g_{||}
\mu_{B}}\right)^{2}\nonumber\\&=&0.579\hspace{1mm} {\rm{K}}^{-1}.
\label{corrcon}
\end{eqnarray}

\begin{figure}[htp]
\resizebox{\hsize}{!}{\includegraphics[clip=true]{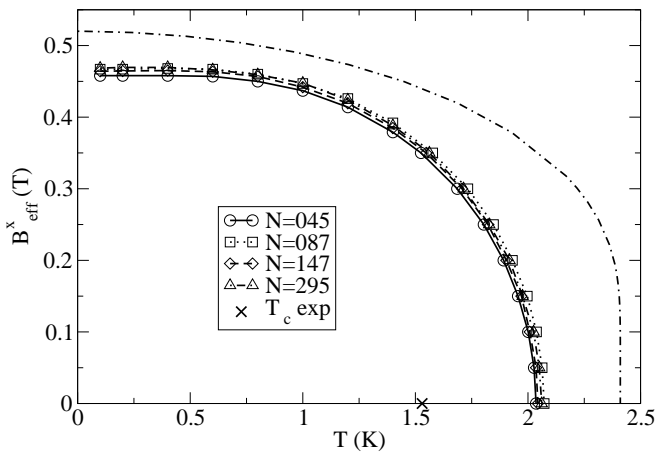}}
\caption{The phase diagram as a function of effective temperature
and effective magnetic field obtained from quantum Monte Carlo
simulations on different lattice sizes. Upper curve is the mean
field solution.} \label{fig:pd}
\end{figure}

The simulation is done over a sequence of spheres with increasing
radius.  We find that the critical curve given by the above
condition converges fairly quickly as the radius is increased. In
the QMC calculation the fluctuations within the sphere are fully
included, and we stress that although we use a mean-field result
for the part of the domain exterior to the sphere, the final
result has converged and should not, in any way, be considered a
mean-field result. One could also argue that the field originating
from dipoles in other domains may affect the critical temperature
in LiHoF$_4$. However, the material forms domains in order to
minimize the energy density of the magnetic field, so this field
should be very small. The same arguments hold also for the case of
a transverse field, and the same condition is used in the presence
of a transverse field.  Fig.~(\ref{fig:pd}) shows the phase
diagram for the mean-field solution as well as for the Monte Carlo
simulation.

The results of Fig.~(\ref{fig:pd}) seem reasonable in the sense
that the QMC solution yields a lower critical temperature than the
mean-field solution. The mean-field solution is within 20\% of the
QMC solution, and it does therefore not suffice to describe the
system at a more precise quantitative level, even for the present
case of a long-range interaction in three dimensions.  The
finite-size effects in Fig.~(\ref{fig:pd}) are quite small already
for moderate system sizes and the result for the largest system
size (N=295 dipoles) is $T_c = 2.03$ K at B$_{x}$=0 and
$(B^{x}_{\rm{eff}})_{C} = 0.47 $ T at the quantum limit. The
critical temperature is still significantly above the experimental
value of $1.53 $ K, but this discrepancy may be attributed to an
additional exchange interaction, which we discuss shortly. The
effective transverse field of 0.47 T corresponds to a physical
transverse field of approximately 3.77 T.  But, as stated earlier,
comparison of the critical field with experimental value at the
quantum limit is possible only when the hyperfine interaction is
included in the model.

\subsection{ Exchange interaction}

A major question at this point is why the critical temperature of
the model is about half a Kelvin higher than the experimental
value. One factor that lowers the critical temperature is the
antiferromagnetic Heisenberg exchange interaction. Since there has
been no direct observation of the strength of the interaction, we
treat it as a free parameter, $J_{\rm{ex}}$, as in
Eq.~(\ref{effham}). First we consider the mean-field solution with
the domain structure. If we assume an exchange interaction of
strength $J_{\rm{ex}}=1.03$ the mean-field critical temperature is
obtained as $T_c=0.2140 \mbox{K} \times (11.271-4\times 1.03)=1.53
$ K, since every spin has four nearest neighbors.

\begin{figure}[htp]
\resizebox{\hsize}{!}{\includegraphics[clip=true]{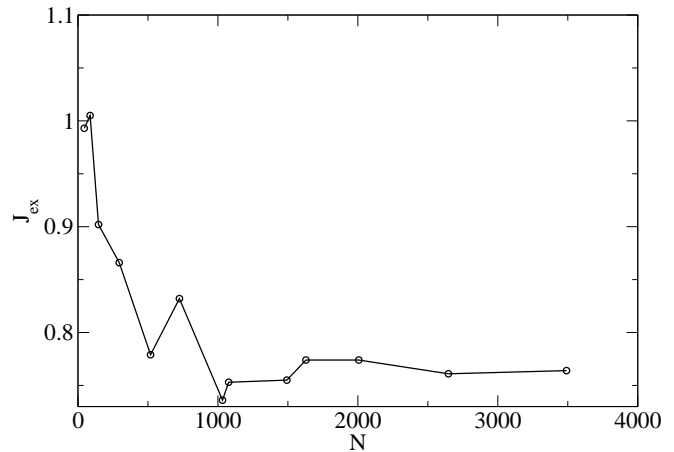}}
\caption{\label{fig:j_exch} Exchange energy needed to tune the
critical temperature. }
\end{figure}

\begin{figure}[htp]
\resizebox{\hsize}{!}{\includegraphics[clip=true]{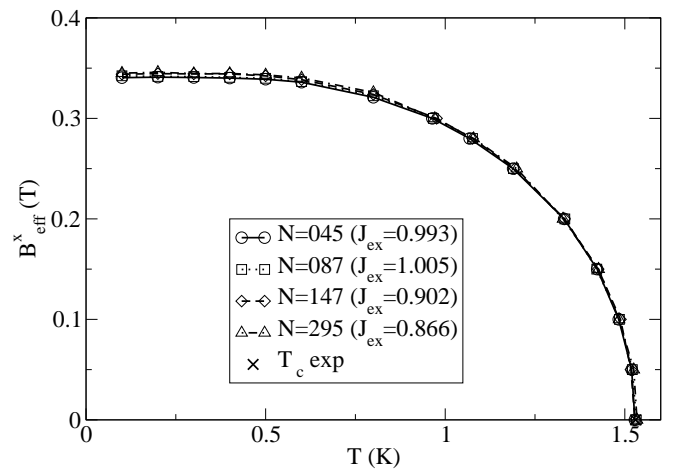}}
\caption{\label{fig:pd_exch} Critical temperature as a function of
effective transverse field, with an exchange interaction}
\end{figure}

Next we perform Monte Carlo simulations, where we tune the value
of the exchange interaction to obtain a critical temperature of
1.53 K. This is done for a sequence of spheres with increasing
radius, and the required exchange parameters are shown in
Fig.~(\ref{fig:j_exch}). The corresponding phase diagram is shown
in Fig.~(\ref{fig:pd_exch}). There is still a strong (and
non-monotone) size dependence in the exchange parameter due to
fluctuations in the number of surface bonds as the radius of the
sphere increases, but the phase diagram shows very little size
dependence. The exchange parameter itself has not converged for
the system sizes we study, but the results for the largest system
sizes averages to about 0.75. This is about 25\% smaller than the
mean-field value, which seems reasonable. A value of
$J_{\rm{ex}}=0.75$ corresponds to an antiferromagnetic exchange
interaction of strength 0.16 K between neighboring spins. This can
be compared to the dipolar interaction between nearest neighbors,
which is ferromagnetic and of strength 0.31 K, or about twice the
exchange interaction. If we instead sum over all bonds connected
to a given site in an ordered domain, the dipolar interaction is
of strength $0.214\times11.271 \mbox{K} = 2.4$ K, while the
exchange energy is of order $0.214\times4\times0.75 \mbox{K} =
0.64$ K. It is difficult to know if this is a reasonable value for
the exchange, and a more stringent test comes when the whole phase
diagram in can be compared to the experimental result. In the next
section, we consider including the hyperfine interaction, which
enables us to compare the phase diagram to experimental data.

The reason for the strong finite-size effect in the exchange
parameter is that the exchange interaction for all broken bonds at
the surface of the sphere is neglected in this calculation. If a
spin is located close to the boundary of the sphere then one, two
or three of its four nearest neighbors may be located outside the
sphere. Even for the largest system size (N=3491) in
Fig.~(\ref{fig:j_exch}), only 80\% of the spins have all four
nearest neighbors inside the sphere. The fraction of spins that
have only one, two or three nearest neighbors inside the sphere
fluctuates very rapidly as the size of the sphere is increased.
However, this is a boundary effect and should disappear as the
system size is increased further.

%\begin{table}[h]
%\begin{ruledtabular}
%\begin{tabular}{c|c}
% System size & Exchange energy $ J_{\rm{ex}}$\\\hline 45 & 0.993\\ 87 &
%1.005\\ 147& 0.902\\ 295 & 0.866\\ 519 & 0.779 \\725 & 0.835\\
% 1033 & 0.736\\ 1493 & 0.758 \\ 2007 & 0.774 \\
%\end{tabular}
%\end{ruledtabular}
%\caption{Exchange energy needed to tune the critical temperature.}
%\label{tab2}
%\end{table}

\section{The hyperfine interaction}

The dynamics of LiHoF$_4$ at low temperatures is complicated by
the hyperfine interaction between the electrons in the Ho$^{3+}$
ion and the Ho-nucleus.\cite{rose96} This interaction is assumed
to be on-site, and its strength is characterized by the hyperfine
constant, A, which is equal to 0.039 K for
LiHoF$_4$.\cite{{menn84},{maga80},{gira01}} At low temperatures,
the hyperfine interaction prefers ordering and the final phase
boundary is a result of the competition between the transverse
field , which tries to destroy the ferromagnetic ordering, and the
hyperfine interaction.

Consider the truncated hamiltonian in Eq.~(\ref{truncate}). Since
the value of the total nuclear angular momentum I in LiHoF$_4$ is
$\frac{7}{2}$, the actual Hilbert space of the electron-nucleus
system for a given site is, in reality, $17 \times 8 =
136$-dimensional. With the inclusion of the hyperfine coupling,
the truncated hamiltonian in Eq.~(\ref{truncate}) should be
written as
\begin{equation}
H_{\rm{hyp}}=\left(V_{C}(\vec{J}) - g_{L}\mu_{B}B_{x}J^{x}\right)
\otimes 1_{N} + A\sum_{\mu=x,y,z}J^{\mu} \otimes I^{\mu},
 \label{nuclearAnonzero}
\end{equation}
where $1_{N}$ is the eight dimensional unit matrix in the nuclear
sector of the Hilbert space.

The hamiltonian in Eq.~(\ref{nuclearAnonzero}) is a
136-dimensional matrix, and in the absence of the hyperfine
interaction, each electronic crystal-field wavefunction is 8-fold
degenerate. The hyperfine interaction breaks this degeneracy to
the order of the strength of A. We adopt the viewpoint that the
transverse magnetic field is renormalized because of the hyperfine
interaction, in a fashion controlled by the temperature. This
renormalization can be extracted by matching the longitudinal
susceptibility of a single Ho$^{3+}$ ion without the presence of
the hyperfine interaction as described below.

\begin{figure}[htp]
\resizebox{\hsize}{!}{\includegraphics[clip=true]{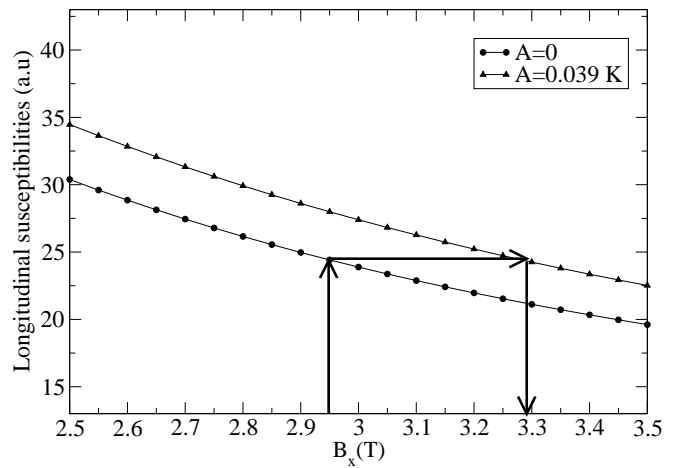}}
\caption{\label{fig:susc} Comparison of the single-ion
longitudinal susceptibilies with and without the hyperfine
interaction.  We make the approximation of assuming that the
increase in longitudinal susceptibility caused by the nuclei can
be modelled by simply renormalizing the transverse field downwards
by the appropriate amount.}
\end{figure}

The susceptibility, $\chi_{zz}$, for a single ion is easy to
define in the 136-dimensional Hilbert space. We diagonalize the
Hamiltonians in Eq.~(\ref{nuclearAnonzero}) with A=0 and A=0.039
K. Let the eigenstates in the 136-dimensional Hilbert space be
denoted by $|\psi_{m}^{0}>$ when A=0 and $|\psi_{m}^{1}>$ when A
is nonzero, and the corresponding energy eigenvalues as
E$_{m}^{0}$ and E$_{m}^{1}$, m=$1\ldots 136$ in both cases. For a
temperature $T=\frac{1}{\beta}$, the longitudinal susceptibility
can be defined as
\begin{eqnarray}
\chi_{zz}^{i}&=&\frac{\beta}{Z_{i}}\sum_{m,n=1\ldots
136}'|\langle\psi_{m}^{i}|J^{z}\otimes
1_{N}|\psi_{n}^{i}\rangle|^{2}e^{-\beta \rm{E}_{m}^{i}}\nonumber\\
& &\frac{-2}{Z_{i}}\sum_{m,n=1\ldots
136}''\frac{|\langle\psi_{m}^{i}|J^{z}\otimes
1_{N}|\psi_{n}^{i}\rangle|^{2}e^{-\beta\rm{E}_{m}^{i}}}{\rm{E}_{m}^{i}-\rm{E}_{n}^{i}},\nonumber\\
& & \label{nucchi}
\end{eqnarray}
where i=0,1. The primed sum is over states degenerate in energy,
while the double-primed sum is over states non-degenerate in
energy. Fig.~(\ref{fig:susc}) shows how the susceptibilities at
the typical temperature T=0.444 K differ for a range of values of
the transverse field depending on whether the hyperfine
interaction is turned on or not.

From the plot of the susceptibilities, it is clear that the
hyperfine interaction renormalizes the magnetic field. This effect
is temperature-dependent, and the shift in susceptibility
decreases as the temperature is increased. For instance, the Ising
mapping and quantum Monte Carlo simulations yield a typical point
on the phase boundary as $(\rm{T_{C}},\rm{B_{x,C}})=(0.444 \rm{K},
2.949 \rm{T})$.  In Fig.~(\ref{fig:susc}) we see that
$\chi^{zz}_{e}(\rm{A}=0,\rm{T}=0.444\rm{K},\rm{B_{x}}=2.949\rm{T})$
equals
$\chi^{zz}_{e}(\rm{A}=0.039\rm{K},\rm{T}=0.444\rm{K},\rm{B_{x}}=3.282\rm{T})$.
Thus, we conclude that the critical transverse field of 2.949 T is
renormalized to the critical transverse field of 3.282 T in the
presence of the hyperfine interaction. This simple renormalization
program can be carried out for all the points on the quantum Monte
Carlo phase boundary. This procedure yields a phase diagram that
shows how the hyperfine interaction affects the phase boundary
significantly in the quantum regime.

\subsection{Magnetic phase diagram from quantum Monte Carlo and hyperfine interaction}
\begin{figure}[htp]
\resizebox{\hsize}{!}{\includegraphics[clip=true]{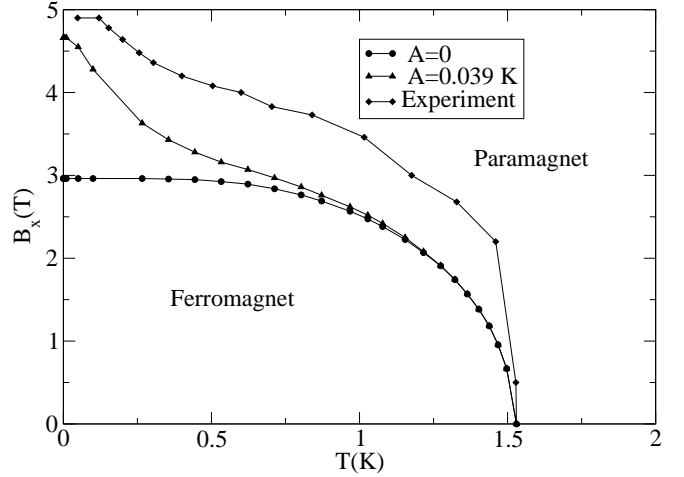}}
\caption{The complete phase diagram of LiHoF$_{4}.$  Experimental
data is from Ref.[\onlinecite{rose96}].}
\label{fig:phasediagram}
\end{figure}

Fig.~(\ref{fig:phasediagram}) shows a comparison of the
theoretically obtained phase diagram with the experimental phase
diagram obtained through susceptibility experiments.\cite{rose96}
We get a quantum critical point of $B_{x,C}=4.66\rm{T}$, which is
approximately 6\% smaller than the experimental value. However,
serious deviations also occur at temperatures higher than
$T\sim1\rm{K}$, where the effect of the nuclear interaction is
negligible. This leads us to conclude that the deviation really
stems from the mapping to the Ising model. The Ising model mapping
is strongly determined by the strength of the degeneracy splitting
in the ground-state doublet as the transverse magnetic field is
turned on, which in turn strongly depends upon the values of the
crystal field parameters. In a subsequent section, the
uncertainties in the crystal field parameters and their effect on
the physics are discussed.

%\begin{figure}[htp]
%\resizebox{\hsize}{!}{\includegraphics[clip=true]{phase_diagram.eps}}
%\caption{The complete phase diagram of LiHoF$_{4}.$  Experimental
%data is from Ref.\onlinecite{rose96}.}
%\label{fig:phasediagram}
%\end{figure}
%SMGTEST

\subsection{Hyperfine interaction and the Ising model}

The actual computation of the longitudinal susceptibility of the
{ion-nucleus} system in the presence of a hyperfine interaction
requires the knowledge of all the eigenstates of the
136-dimensional Hilbert space. However, the physics of why the
hyperfine interaction enhances the longitudinal susceptibility can
be understood by considering a toy system consisting of a single
spin-$\frac{1}{2}$ electron coupled to a spin-$\frac{1}{2}$
nucleus through a hyperfine interaction. It shall be seen that the
enhancement of susceptibility is really a subtle effect captured
in second-order perturbation theory.

Let us consider a single spin-$\frac{1}{2}$ electron placed in a
transverse magnetic field $h_{x}$ (we assume that the nucleus does
not couple to the magnetic field: this is a reasonable assumption
since the nuclear g-factor in LiHoF$_{4}$ is approximately 1000
times smaller than the electronic Land\'{e} g-factor). Then the
hamiltonian is
\begin{equation}
H=-h_{x}\sigma^{x}_{e}\otimes 1_{N}. \label{Isingnuclearfirst}
\end{equation}
The subscripts $e$ and $N$ refer to the electronic and nuclear
degrees of freedom, respectively. The Hilbert space is $((2\times
\frac{1}{2} + 1)(2\times\frac{1}{2} + 1)=)4$-dimensional, and the
transverse field splits the Hilbert space in two multiplets, each
of which is twofold degenerate. The ground electronic multiplet
consists of the states
$|\rightarrow\rangle_{e}\otimes|\uparrow\rangle_{N}$ and
$|\rightarrow\rangle_{e}\otimes|\downarrow\rangle_{N}$, while the
excited electronic multiplet consists of the states
$|\leftarrow\rangle_{e}\otimes|\uparrow\rangle_{N}$ and
$|\leftarrow\rangle_{e}\otimes|\downarrow\rangle_{N}$, the
difference in energy between the two multiplets being $2h_{x}$.
The longitudinal susceptibility can be easily evaluated using the
expression in Eq.~(\ref{nucchi}) (with suitable modifications for
the toy system), and we find
\begin{equation}
\chi_{zz}=\frac{1}{h_{x}} \label{Isingnuclearsecond}
\end{equation}
at $T=0$.

If the hyperfine interaction is turned on, the degeneracies
between the states in the nuclear sector is lifted in each
multiplet, and this changes the susceptibilities. Let the
perturbing hyperfine interaction be written as
\begin{equation}
V_{\rm{hyp}}=A_{\parallel}\sigma^{z}_{e}\sigma^{z}_{N} +
A_{\perp}\sigma^{x}_{e}\sigma^{x}_{N}. \label{Isingnuclearthird}
\end{equation}
Even though the hyperfine strengths are isotropic in the physical
system, introducing anisotropy helps us to understand the roles
played by the longitudinal and transverse components of the
hyperfine interaction transparently. The special case of isotropy,
$A_{\parallel}=A_{\perp}$, will be considered at the end.

In first order degenerate perturbation theory, the electronic
states are all polarized in the x-direction, and only $A_{\perp}$
has any non-zero matrix element within the same degenerate
electronic multiplet. Thus $A_{\parallel}$ drops out of the
physics in first-order perturbation theory.

%SMGTEST

\begin{figure}[htp]
\resizebox{\hsize}{!}{\includegraphics[clip=true]{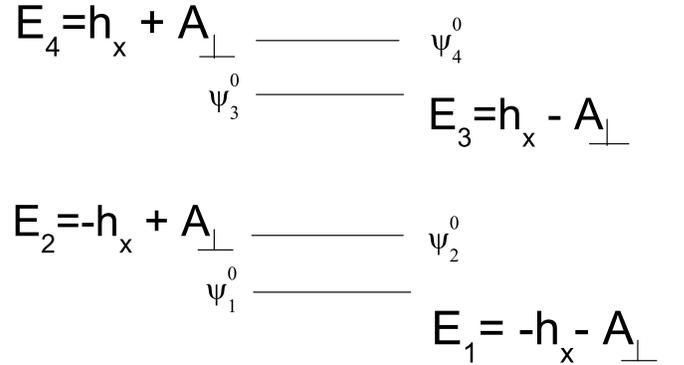}}
\caption{ Ion energy spectrum (for a spin 1/2 nucleus) with the
hyperfine interaction treated in first order perturbation theory}
\label{fig:first-order}
\end{figure}
In Fig.~(\ref{fig:first-order}), the states in the four
dimensional Hilbert space are
$|\psi_{1}^{0}\rangle=|\rightarrow\rangle_{e}\otimes|\leftarrow\rangle_{N}$,
$|\psi_{2}^{0}\rangle=|\rightarrow\rangle_{e}\otimes|\rightarrow\rangle_{N}$,
$|\psi_{3}^{0}\rangle=|\leftarrow\rangle_{e}\otimes|\rightarrow\rangle_{N}$
and
$|\psi_{4}^{0}\rangle=|\leftarrow\rangle_{e}\otimes|\leftarrow\rangle_{N}$.
The zero temperature susceptibility is now given by
\begin{equation}
\chi_{zz}=\frac{1}{h_{x}+\mid A_{\perp}\mid}
\label{Isingnuclearfourth}.
\end{equation}

Thus it is obvious that the longitudinal susceptibility always
decreases in first-order perturbation theory. The transverse
component of the hyperfine interaction always acts as an extra
transverse field, which implies that the hyperfine interaction
should lower the critical field, at least in first-order
perturbation theory.

However, if we consider second-order perturbation theory, we can
show that, when the hyperfine interaction is \emph{isotropic}, the
net effect is actually an increase in the longitudinal
susceptibility at $T=0$. In second-order, only $A_{\parallel}$
contributes to further splitting of the energy spectrum. More
importantly, the ground-state
$|\rightarrow\rangle_{e}\otimes|\leftarrow\rangle_{N}$ is mixed
with the state
$|\leftarrow\rangle_{e}\otimes|\rightarrow\rangle_{N}$, and the
state $|\rightarrow\rangle_{e}\otimes|\rightarrow\rangle_{N}$ is
mixed with the state
$|\leftarrow\rangle_{e}\otimes|\leftarrow\rangle_{N}$ both with an
amplitude given by $\varepsilon=\frac{A_{\parallel}}{h_{x}}$.
Thus, up to second-order perturbation theory we have the following
spectrum:
\begin{eqnarray}
|\psi_{1}\rangle&=&|\psi^{0}_{1}\rangle -
\frac{A_{\parallel}}{h_{x}}|\psi^{0}_{3}\rangle\nonumber\\
|\psi_{2}\rangle&=&|\psi^{0}_{2}\rangle -
\frac{A_{\parallel}}{h_{x}}|\psi^{0}_{4}\rangle\nonumber\\
|\psi_{3}\rangle&=&|\psi^{0}_{3}\rangle +
\frac{A_{\parallel}}{h_{x}}|\psi^{0}_{1}\rangle\nonumber\\
|\psi_{4}\rangle&=&|\psi^{0}_{4}\rangle +
\frac{A_{\parallel}}{h_{x}}|\psi^{0}_{2}\rangle .
\label{secondorderhyperfinestates}
\end{eqnarray}
The energies of the states up to second-order in perturbation
theory are given by
\begin{eqnarray}
E_{1}&=&-h_{x}-A_{\perp}-\frac{A_{\parallel}^{2}}{2h_{x}}\nonumber\\
E_{2}&=&-h_{x}+A_{\perp}-\frac{A_{\parallel}^{2}}{2h_{x}}\nonumber\\
E_{3}&=&+h_{x}-A_{\perp}+\frac{A_{\parallel}^{2}}{2h_{x}}\nonumber\\
E_{4}&=&+h_{x}+A_{\perp}+\frac{A_{\parallel}^{2}}{2h_{x}} .
\label{secondorderhyperfineenergies}
\end{eqnarray}
This gives rise to a non-zero matrix element of $\sigma^{z}_{e}$
between $|\psi_{1}\rangle$ and $|\psi_{2}\rangle$, and this term
serves to cancel the decrease in susceptibility because of
first-order contribution from $A_{\perp}$. The exact expression
for the longitudinal susceptibility at $T=0$ is given by
\begin{equation}
\chi_{zz}=\frac{(1-(\frac{A_{\parallel}}{2h_{x}})^{2})^{2}}{h_{x}+A_{\perp}+\frac{A_{\parallel}^{2}}{2h_{x}}}
+\frac{A_{\parallel}^{2}}{2h_{x}A_{\perp}} .
\label{Isingnuclearfifth}
\end{equation}
The denominator in the first term in the RHS of
Eq.~(\ref{Isingnuclearfifth}) can be easily expanded in terms of
the ratios $\frac{A_{\perp}}{h_{x}}$ and
$\frac{A_{\parallel}^{2}}{2h_{x}^{2}}$, and the susceptibility
expression turns out to be $(h_{x}\gg A_{\parallel},A_{\perp})$
\begin{eqnarray}
\chi_{zz}&=&\frac{1}{h_{x}}+(\frac{A_{\parallel}^{2}}{A_{\perp}} -
A_{\perp})\frac{1}{h_{x}^{2}}\nonumber\\
& &+ O(\frac{1}{h_{x}^{3}}) + \ldots
 \label{Isingnuclearsixth}
\end{eqnarray}

The crux of the increase in susceptibility lies in the fact that
the term proportional to $\frac{1}{h_{x}^{2}}$ is positive when
$A_{\parallel}\geq A_{\perp}$. The increase in susceptibility is
not a generic feature of the \emph{ion-nuclear} interaction for
all values of $A_{\parallel}$ and $A_{\perp}$, but is present when
the interaction is isotropic, as in LiHoF$_{4}$. The transverse
component always drives the susceptibility lower, but the
longitudinal component can compete and win for a certain range of
values, and the change in susceptibility is quadratic in
$A_{\parallel}$. We emphasize that this positive contribution
arises from the non-zero matrix element between two states that
were degenerate in the absence of the hyperfine interaction.

This schematic problem of a spin-$\frac{1}{2}$ electron coupled to
a spin-$\frac{1}{2}$ nucleus through a hyperfine interaction
serves nicely to illustrate the physics behind an increase in the
longitudinal susceptibility of the electrons due to the hyperfine
interaction. In LiHoF$_4$, every electronic state is split into
eight nuclear states when the hyperfine interaction is turned on.
There is no matrix element of $J_{z}$ within the same electronic
multiplet when $A_{\parallel}=0$. When $A_{\parallel}\neq 0$, a
non-zero matrix element of $J_{z}$ develops between two lowest
nuclear states in the ground electronic multiplet, the square of
the matrix element varying quadratically as $A_{\parallel}$.
Exactly as outlined in the schematic problem, this matrix element
causes an increase in the longitudinal susceptibility of the
electron.

\subsection{Projecting the hyperfine interaction}

The hyperfine interaction involves the electronic angular momentum
$J^{\mu}$. Following the mapping to the Ising subspace in the
previous sections, we can map these degrees of freedom to
effective Ising degrees of freedom. This gives rise to a hyperfine
interaction term that operates in an effectively 16(=2$\times$
8)-dimensional Hilbert space. However, this gives rise to a
hyperfine interaction whose strength depends on the magnetic
field. Using the mapping in Eq.~(\ref{splitting}),
Eq.~(\ref{map1}) and the hyperfine interaction in
Eq.~(\ref{nuclearAnonzero}), the effective single-site hamiltonian
now becomes
\begin{equation}
H=-\frac{\Delta(B_{x})}{2}\sigma^{x}_{e}\otimes 1_{N} +
AC_{\alpha\beta}(B_{x})\sigma^{\alpha}\otimes I^{\beta} .
\label{hyperfineprojection}
\end{equation}

Using the projected hamiltonian in
Eq.~(\ref{hyperfineprojection}), we can naturally understand why
the longitudinal component of the hyperfine interaction wins over
the transverse one. The longitudinal component is proportional to
$C_{zz}(B_{x})$, which is significantly larger than all the other
components at the magnetic field regime of interest. The analysis
of the previous section remains valid in a 16-dimensional Hilbert
space, and the susceptibility is always enhanced due to the
contribution from the longitudinal component of the hyperfine
interaction $A_{\parallel}=AC_{zz}(B_{x})$.

\subsection{Uncertainties in the Crystal Field Parameters}

The quantitative details of the LiHoF$_{4}$ depend sensitively on
the values of the various crystal field parameters (CFP) used in
constructing the crystal field hamiltonian (Please see Appendix
for details). For example, we have found that the value of the
critical transverse field, $B_{x,C}$, varies by as large as 25\%
when the CFP $B_{2}^{0}$ is varied by 10\%. However, the
parameters cannot be determined directly by experiments, but are
used as fitting parameters to fit theoretical calculations to
experimental data. There have been attempts to determine the CFP-s
by fitting to spectroscopic data\cite{chri79} or to susceptibility
measurements.\cite{hans75} However, we have found that the results
of theoretical calculations become more and more sensitive to the
values of CFP-s as the transverse field is increased. As we have
shown in earlier sections, the most important physical quantity
that determines the phase diagram is the splitting between the two
lowest states, $\Delta(B_{x})$, that smoothly and monotonically
increases with the transverse field. Thus we propose a
spectroscopic experiment in the presence of a transverse field to
determine $\Delta(B_{x})$. However, the quantum regime of the
phase diagram is complicated due to the presence of hyperfine
interaction. The effect of the hyperfine interaction is also
extremely sensitive to the values of the CFP-s, and this effect is
very difficult to calculate theoretically. Thus we propose the
experiment be carried out in a regime of the phase diagram where
the quantum fluctuations due to the hyperfine interaction are
negligible, but the fluctuations caused by the transverse magnetic
field are still significant. Thus we propose an experiment to
determine $\Delta(B_{x})$ in the magnetic field range 2.0 - 3.0 T.
In this regime, our theoretical calculations show that the
splitting energy varies between $\sim$ 1.6 K - 3.0 K. This
corresponds to a microwave frequency range of $\sim$ 20 - 70 GHz.
A spectroscopic experiment that determines $\Delta(B_{x})$
accurately is the most important ingredient needed to determine
the phase diagram accurately. Even though a single spectroscopic
experiment is not enough to determine the values of all the
crystal field parameters uniquely, we have found that the only
other important parameter in the effective Ising hamiltonian,
$C_{zz}(B_{x})$, is extremely robust even to large changes in the
crystal field parameters. Thus, an accurate determination of
$\Delta(B_{x})$ is enough to compute the phase diagram from the
effective Ising model.

\section{Summary and conclusions}

It has been postulated that LiHoF$_4$ is an example of an
Ising-like system, and when the sample is placed in a magnetic
field transverse to the Ising direction, it is an example of a
quantum Ising system with magnetic dipole interaction being the
ordering interaction. However, the quantum Ising model,
Eq.~(\ref{dipole}) has never been used to obtain physical results
for the system. We derive the physical Ising model by a
non-perturbative mapping to the Hilbert space spanned by the
ground-state doublet. We then do a quantum Monte Carlo simulation
to obtain the phase diagram, also incorporating the domain
structure in the process. This is a step beyond mean-field theory
and the calculation is sufficiently accurate that uncertainty in
the predicted phase diagram is now limited by uncertainties in the
crystal field parameters. As a spin-off, we are able to compute
the phenomenological exchange interaction parameter that modifies
the phase diagram considerably in the classical regime. The
hyperfine interaction poses considerable problems in comparing the
phase diagram obtained from the quantum Monte Carlo simulations to
the experimental data. We have made the approximation that the
effects of the hyperfine interaction can be completely recovered
through a renormalization of the magnetic field.

LiHoF$_4$ is a material in which the magnetic quantum and
classical phase transitions can be controlled with great
precision. Neutron scattering studies have been done on LiHoF$_4$
to obtain the spin-wave excitations in the system. By randomly
replacing the magnetic Ho$^{3+}$ with non-magnetic Y$^{3+}$ ions,
spin-glass behaviour has been observed. We believe that the
existence of an Ising model that faithfully reproduces the physics
in both the classical and the quantum regimes, will facilitate the
investigations of the interesting properties of LiHoF$_4$
considerably.

\begin{acknowledgments}
We thank Jens Jensen for many helpful discussions and for
supplying us with a copy of Ref.[6]. PH acknowledges support by
the Swedish Research Council and the G\"oran Gustafsson
foundation. SMG and PBC were supported by NSF DMR-0342157. AWS was
supported by the Academy of Finland, Project No. 26175.
\end{acknowledgments}

\subsection{Appendix: The Crystal Field Hamiltonian} In LiHoF$_4$, the
Ho$^{3+}$ ions have an unfilled shell 4$f$ with 10 electrons. The
Hunds' rules dictate that the ground configuration of a single
Ho$^{3+}$ ion should be $^{5}I_{8}(S=2, L=6, J=8)$. If there were
no interactions with the neighboring ions, the ground state of a
single ion will be 17-fold degenerate. However, the Coulomb
interactions with the neighboring ions gives rise to an electric
field that lifts this degeneracy. In general, this electric field
depends strongly on the spatial symmetry of the crystals. In the
simplest scenario, each ion is regarded as a point charge, and the
spatial overlap of the wave function of an ion with its
neighboring ion is neglected. This is the point-charge model for
calculating crystal fields.

The derivation of the crystal field electrostatic potential takes
into account the lattice symmetry, and is most simply expressed in
terms of the well-known spherical harmonics. We shall not discuss
the details of the derivation here, but instead refer the reader
to the article by M. T. Hutchings,\cite{hutc64} and references
therein.  If the electrostatic potential at a point
$\vec{r}=(x,y,z)$ near the ion of interest is denoted by
$V(x,y,z)$, the crystal field hamiltonian can simply be written as
\begin{equation}
V_{C}=-|e|\sum_{i=1}^{n_{e}}V(\hat{x}_{i},\hat{y}_{i},\hat{z}_{i}).
\label{potential}
\end{equation}
In Eq.~(\ref{potential}), $n_{e}$ denote the number of electrons
in the unfilled shell (10 in case of LiHoF$_{4}$), and
$(\hat{x}_{i},\hat{y}_{i},\hat{z}_{i})$ are the quantum position
operators of the $i$th electron.

Expressed in terms of the individual electron operators the
crystal field hamiltonian still presents a formidable problem, but
it is much simplified if we consider the outer shell electrons in
an appropriate coupling scheme. In case of 4$f$ electrons, the
ionic states are usually expressed in terms of the quantum numbers
$|L,S,J,J_{z}\rangle$. If we restrict ourselves to a single value
of J (neglecting configuration mixing), the hamiltonian in
Eq.~(\ref{potential}) can be expressed in terms of \emph{Operator
Equivalents}.\cite{{stev52},{blea53}} This eliminates the need to
use a Slater determinant consisting of single-electron wave
functions to evaluate the matrix elements of $V_{C}$.

The Operator Equivalents are operators built out of the $\vec{J}$
operators that act on the $(2J + 1)$-dimensional space determined
by the value of $J$. However, they act only on the angular part of
the wave function of the coupled system, and the matrix elements
of the radial part of the wave function are usually incorporated
as fitting parameters.

The number of operators needed to completely determine the
hamiltonian in Eq.~(\ref{potential}) and the rules for deriving
them, depend on the symmetry of the crystal and the ground state
configuration of the ion. These rules are clearly explained by K.
W. H. Stevens.\cite{stev52} Here we shall just list the operators
that have non-zero matrix elements in the configuration
$^{5}I_{8}$ of the Ho$^{3+}$ ion in LiHoF$_{4}$.

The operators are usually denoted by two indices, $l$ and $m$,
corresponding to the same indices on the spherical harmonics they
are equivalent to. There is an additional index, $C(S)$, for
non-zero $m$, corresponding to a symmetric(anti-symmetric)
combination of the underlying spherical harmonics $Y_{l}^{m}$ and
$Y_{l}^{-m}$.

In case of LiHoF$_4$, the relevant crystal field operators are:
\begin{eqnarray}
O_{2}^{0}&=&3J_{z}^{2}-J(J+1)\nonumber\\
O_{4}^{0}&=&35J_{z}^{4}-30J(J+1)J_{z}^{2} + 25J_{z}^{2}\nonumber\\
& &- 6J(J+1)
+ 3J^{2}(J+1)^{2}\nonumber\\
O_{4}^{4}(C)&=&\frac{1}{2}(J_{+}^{4} + J_{-}^{4})\nonumber\\
O_{6}^{0}&=&231J_{z}^{6}-315J(J+1)J_{z}^{4} +
735J_{z}^{4}\nonumber\\
& &+105J^{2}(J+1)^{2}J_{z}^{2}-525J(J+1)J_{z}^{2} +
294J_{z}^{2}\nonumber\\
& &-5J^{3}(J+1)^{3} +
40J^{2}(J+1)^{2} -60J(J+1)\nonumber\\
O_{6}^{4}(C)&=&\frac{1}{4}(J_{+}^{4} +
J_{-}^{4})(11J_{z}^{2}-J(J+1)-38) + h.c.\nonumber\\
O_{6}^{4}(S)&=&\frac{1}{4i}(J_{+}^{4} -
J_{-}^{4})(11J_{z}^{2}-J(J+1)-38) + h.c.\nonumber\\
\label{CFoperators}
\end{eqnarray}
where $J_{+}=J_{x}+iJ_{y}$ and $J_{-}=J_{x}-iJ_{y}$.

Using these operators, the crystal field hamiltonian $V_{C}$ can
be written as:
\begin{eqnarray}
V_{C}&=&B_{2}^{0}O_{2}^{0} + B_{4}^{0}O_{4}^{0} +
B_{6}^{0}O_{6}^{0} + B_{4}^{4}(C)O_{4}^{4}(C)\nonumber\\
& &+ B_{6}^{4}(C)O_{6}^{4}(C) + B_{6}^{4}(S)O_{6}^{4}(S).
\label{CFhamiltonian}
\end{eqnarray}
The radial matrix elements of the electrostatic potential is
extremely difficult to compute accurately even in a point-charge
model. They are, therefore, incorporated within the constants
$B_{l}^{m}$, known as crystal field parameters(CFP). The CFPs are
generally used as fitting parameters. In LiHoF$_{4}$, for example,
they are used to fit the crystal field spectrum to observed
spectroscopic data,\cite{{chri79},{maga80},{sala97},{agla93}} and
to susceptibility measurements.\cite{hans75}

In all our calculations, we use the CFPs proposed by R{\o}nnow
\emph{et al.}\cite{jens01} Their values (in K) are listed below:
\begin{eqnarray}
B_{2}^{0}&=&-0.696\nonumber\\
B_{4}^{0}&=&4.06\times10^{-3}\nonumber\\
B_{6}^{0}&=&4.64\times10^{-6}\nonumber\\
B_{4}^{4}(C)&=&0.0418\nonumber\\
B_{6}^{4}(C)&=&8.12\times10^{-4}\nonumber\\
B_{6}^{4}(S)&=&1.137\times10^{-4}.
\end{eqnarray}

These values of CFPs were obtained\cite{jens01} by fitting the
results of RPA spin-wave dynamics calculations to observed neutron
scattering data, as well as to the two lowest energy levels of the
crystal field spectrum, as observed in spectroscopic
measurements.\cite{gife78} However, there are no estimates of the
accuracies to which these parameters are known. There was an
earlier attempt to determine the CFPs by fitting to susceptibility
data,\cite{hans75} but there were very large error bars. Another
attempt was made to determine the CFPs by fitting to spectroscopic
measurements,\cite{chri79} but an incorrect symmetry($D_{2d}$) of
the crystal was used in the theoretical calculations, instead of
the correct one, $S_{4}$.

\bibliography{bib}

\end{document}